\documentclass[reprint,aps,prl,amsmath,amssymb,showpacs,nofootinbib]{revtex4-1}

\usepackage[pdftex]{graphicx}
\usepackage{xcolor}
\graphicspath{ {figs/} }

\usepackage{hyperref}
\hypersetup{
	breaklinks=true,
	colorlinks=true,
	linkcolor=blue,
	citecolor=purple,
	pdfstartview={FitH},
	pdfview={FitH 0},
	pdfauthor={Jeffrey B. Parker},
	pdftitle={Topological gaseous plasmon polariton in realistic plasma},
 }

\usepackage{bm}
\usepackage{eucal}		
\usepackage{mycommands}   

\providecommand{\eqnref}{}
\renewcommand{\eqnref}[1]{Eq.~\eqref{#1}}

\newcommand{\vk}{\v{k}}
\usepackage{physics}

\begin{document}

\title{Topological Gaseous Plasmon Polariton in Realistic Plasma}

\author{Jeffrey B.~Parker}
\email{jbparker3@wisc.edu}
\affiliation{Lawrence Livermore National Laboratory, Livermore, California 94550, USA}

\author{J.~B.~Marston}
\affiliation{Brown Theoretical Physics Center and Department of Physics, Brown University, Providence, Rhode Island 02912-1843, USA}

\author{Steven M.~Tobias}
\affiliation{Department of Applied Mathematics, University of Leeds, Leeds, LS2 9JT, United Kingdom}

\author{Ziyan Zhu}
\affiliation{Department of Physics, Harvard University, Cambridge, Massachusetts 02138, USA}

\begin{abstract}
Nontrivial topology in bulk matter has been linked with the existence of topologically protected interfacial states.  We show that a gaseous plasmon polariton (GPP), an electromagnetic surface wave existing at the boundary of magnetized plasma and vacuum, has a topological origin that arises from the nontrivial topology of magnetized plasma.  Because a gaseous plasma cannot sustain a sharp interface with discontinuous density, one must consider a gradual density falloff with scale length comparable to or longer than the wavelength of the wave.  We show that the GPP may be found within a gapped spectrum in present-day laboratory devices, suggesting that platforms are currently available for experimental investigation of topological wave physics in plasmas. 
\end{abstract}

\maketitle

Edge states arising from topologically nontrivial bulk matter have attracted significant recent attention.  For example, topological insulators and the quantum Hall states are now understood as manifestations of topology~\cite{hasan:2010,qi:2011}, and similar reasoning has been applied to a diverse range of other physical systems~\cite{martin:2008}.  Edge states have garnered intense practical interest due to topological protection and the prospect for robust, undirectional propagation with reduced losses to scattering from defects.  In analogy to the systems in the quantum mechanical regime, classical systems, including photonics~\cite{wang:2009,feng:2011,plotnik:2014,skirlo:2014,lu:2014,skirlo:2015,gangaraj:2018}, acoustics~\cite{peano:2015,yang:2015,he:2016}, mechanical systems~\cite{nash:2015,huber:2016}, as well as continuum fluids~\cite{silveirinha:2015,delplace:2017,shankar:2017,perrot:2019}, can exhibit topological quantization as well as edge states between topologically distinct states of matter.

Plasmas support rich wave physics, especially in the presence of a magnetic field, multiple species, kinetic distributions, and inhomogeneity.  Analysis of band structure and wave dispersion properties has been a cornerstone in the understanding of plasma waves, leading to important practical applications such as current drive and heating for fusion devices.  Yet the topological characterization of plasma band structure, and its consequences for edge states, has not been fully appreciated.  One recent work has proposed that the reversed-shear Alfv\'{e}n eigenmode observed in tokamaks arises from the nontrivial topology associated with magnetic shear and the topological phase transition across a zero-shear layer \cite{parker:2019:rsae}.

Here we consider one of the simplest plasmas, a dilute gas of ions and electrons in a magnetic field.  We investigate the topological gaseous plasmon polariton (GPP), an electromagnetic surface wave that arises due to nontrivial topology of a magnetized plasma.  The applied magnetic field breaks time-reversal symmetry, and therefore the topology is analogous to that of the integer quantum Hall effect \cite{thouless:1982} or Kelvin waves \cite{delplace:2017}.  While other surface waves in inhomogeneous or bounded plasmas have been investigated previously \cite{trivelpiece:1959,breizman:2000}, topological aspects have not been considered.  The GPP initially appears from similar mathematical structure as the surface magnetoplasmon polariton occurring at the surface of metals or semiconductors \cite{brion:1972,bin:2015}.  However, the internal structures of metals and plasmas are quite different, and quantum corrections are unlikely to play an important role in the gaseous plasma considered here.  Another critical difference is that gaseous plasmas, unlike metals and semiconductors, cannot sustain sharp interfaces where the density jumps essentially discontinuously.  The spatial variation of the plasma density introduces additional physics such as a changing upper hybrid frequency, and the character of the local dispersion relation may shift across the plasma.  The question of whether or not a plasma can support the GPP when the density varies over a length scale larger than a wavelength has not yet been addressed.  Potential uses of surface waves like the GPP include surface-mode-sustained plasma discharges \cite{ferreira:1981,aliev:book}.

We consider a plasma with realistic density profile and demonstrate the GPP can be supported, and furthermore we show that parameter regimes in which the GPP is accessible may be attained in currently existing laboratory devices.  Our results motivate experiments to probe many of the open issues regarding topological waves in plasmas, such as to what extent they exhibit topological protection, and how nonlinearities affect their behavior.

We adopt the cold-plasma model of a magnetized, stationary plasma, appropriate for light waves when the electron thermal speed is much less than the speed of light.  We assume a high-frequency regime and retain only the electron motion, treating ions as an immobile neutralizing background.  Electron collisions are neglected for the dilute plasmas considered here because the collision frequency is orders of magnitude smaller than the wave frequency of interest.  The linearized equations of motion for an infinite homogeneous plasma are \cite{stix:book}
	\begin{subequations}
	\label{homogeneous_cold_plasma_equations}
	\begin{align}
		\pd{\v{v}}{t} &= -\frac{e}{m_e} (\v{E} + \v{v} \times \v{B}_0), \\
		\pd{\v{E}}{t} &= c^2 \nabla \times \v{B} + \frac{e n_e}{\e_0} \v{v}, \\
		\pd{\v{B}}{t} &= -\nabla \times \v{E},
	\end{align}
	\end{subequations}
where $\v{v}$ is the electron fluid velocity, $\v{E}$ the electric field, $\v{B}_0 = B_0 \unit{z}$ the background magnetic field, $\v{B}$ the perturbation magnetic field, $e$ the elementary charge, $n_e$ the background electron density, $m_e$ the electron mass, $c$ the speed of light, and $\e_0$ the permittivity of free space.  It is convenient to work in nondimensionalized units in which time is normalized to $\w_p^{-1}$, where $\w_{p} = (n_e e^2 / m_e \e_0)^{1/2}$ is the plasma frequency, length to $c/\w_p$, velocity to $e \ol{E} / m_e \w_p$, electric field to $\ol{E}$, and magnetic field to $\ol{E}/c$, where $\ol{E}$ is some reference electric field.   Then the only parameter is $\s = \sign(B_0) \W_e / \w_{p}$, where $\W_e = |eB_0/m_e|$ is the electron cyclotron frequency.  Upon letting $\partial/\partial t \to -i \w$ and $\nabla \to i\v{k}$, we obtain the eigenvalue equation $H \ket{f} = \w \ket{f}$, where $\ket{f} = \begin{bmatrix} \v{v} & \v{E} & \v{B} \end{bmatrix}$ is a 9-element vector and $H$ is a $9 \times 9$ Hermitian matrix corresponding to the linear operator, which plays the role of an effective Hamiltonian.  We work in Cartesian coordinates.  The matrix $H$ is written out explicitly in the Supplemental Material \cite{suppmaterial}.

\begin{figure}
		\includegraphics{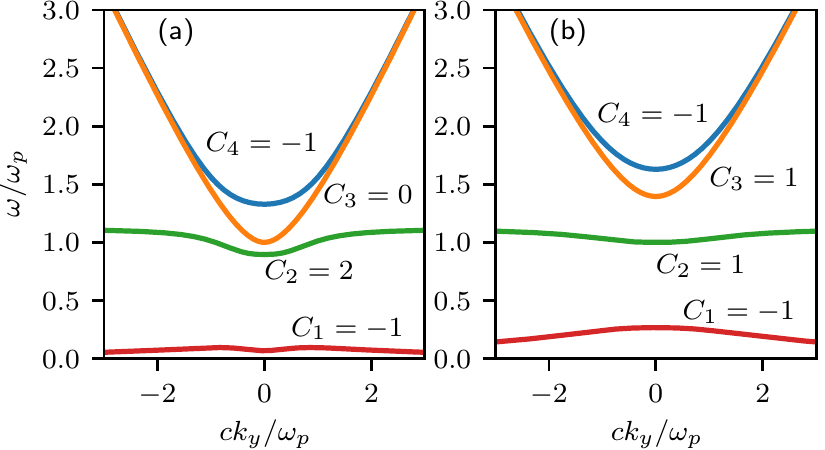}
		\caption{Spectrum of a magnetized, homogeneous cold plasma as a function of $k_y$  ($k_x$ set to zero, but the system is isotropic in the $xy$ plane), where only electron motion is retained.  Here, $\s = 0.5$.  (a) $c k_z / \w_p = 0.4$ ($k_z < k_z^*$).  (b) $c k_z / \w_p = 1.1$ ($k_z > k_z^*$).  Also shown are the Chern numbers of the positive-frequency bands, computed with the use of the discreteness regularization.}
		\label{fig:homog_plasma_spectrum}
	\end{figure}

We allow for an arbitrary propagation angle with respect to the magnetic field.  We fix $k_z$ and consider a parameter space $\vk_\perp = (k_x, k_y)$.  This problem is isotropic in the plane perpendicular to the magnetic field.  For each $\vk_\perp$, there are 9 solutions for the eigenvalues $\w_n$, for $n=-4, -3, \ldots, 4$, which we order by ascending frequency, and $\w_{-n} = -\w_n$.  The corresponding eigenfunctions are denoted $\ket{n}$.  Except for certain values of $k_z$ and $\s$, the eigenvalues are nondegenerate.  The band structure is shown in Fig.~\ref{fig:homog_plasma_spectrum}.

When the eigenvalues are nondegenerate, frequency band $n$ may be characterized by a Chern number $C_n = (2\pi)^{-1} \int d\vk_\perp \, F_n(\vk_\perp)$, where the Berry curvature $F_n(\vk)$ of the band at a given $\vk$ is given by
	\begin{widetext}
	\begin{equation}
		F_n (\vk) = i \sum_{m \neq n} \frac{\matrixelement**{n}{\displaystyle \pd{H}{k_x}}{m}  \matrixelement**{m}{\displaystyle \pd{H}{k_y}}{n} - \matrixelement**{m}{\displaystyle \pd{H}{k_x}}{n}  \matrixelement**{n}{\displaystyle \pd{H}{k_y}}{m}}{(\omega_n - \omega_m)^2}.
		\label{berry_curvature_gaugeinvariant}
	\end{equation}
	\end{widetext}

The Chern numbers $C_1, C_3, C_4$ are integer valued, but $C_2$ takes noninteger values when using the linear operator in \eqnref{homogeneous_cold_plasma_equations}.  The issue of a ``noninteger Chern number'' stems from a lack of insufficient smoothness at small scales of the linear operator $H$, which leads to the inability to compactify the infinite $\vk$ plane \cite{silveirinha:2015}.  If $H$ falls off sufficiently rapidly, one can map the $\vk$ plane into the Riemann sphere, which is compact.

An integer $C_2$ can be restored through regularization of $H$.  Regularization of continuous electromagnetic media based on the notion of material discreteness has been addressed previously~\cite{silveirinha:2015,hanson:2016arxiv}, although other means of compactifying continuum fluids have been discussed~\cite{tauber:2019,souslov:2019}.  The plasma ceases to look like a continuous medium at sufficiently small length scales, which the fluid model does not take into account.  To model this physical discreteness, the regularization suppresses the plasma response at small scales.  In the Fourier representation, the nondimensionalized electron equation of motion is modified to be $\partial_t \v{v} = -r(\vk) \v{E} - \s \v{v} \times \unit{z}$, where $r$ becomes small at large wave vectors. For instance, we can take $r(\vk) = (1 + |\vk_\perp|^2/k_c^2)^{-1}$ for some cutoff wave number $k_c$.  This modification functionally alters the plasma frequency to become small at small length scales.  To preserve Hermiticity, we also modify the nondimensionalized Ampere--Maxwell equation to be $\partial_t \v{E} = \nabla \times \v{B} + r(\vk) \v{v}$.  

Using the discreteness regularization, the Chern numbers are integer valued and are independent of the form of $r(\vk)$ as long as it decays sufficiently rapidly.  The regularization removes noninteger contributions from infinite $\vk$ while retaining the integer-valued contribution from finite $\vk$.  A key point is that $k_c$ may be taken to be arbitrarily large, such that the physical effect of the regularization at length scales of interest can be made arbitrarily small.  Additional details can be found in the Supplemental Material \cite{suppmaterial}.

We consider $k_z > 0$.  Because bands 2 and 3 touch at $k_z = k_z^*$, where $(ck_z^*/\w_p)^2 = \s / (1 + \s)$, there are two distinct regimes: $k_z < k_z^*$ and $k_z > k_z^*$.  At $k_z < k_z^*$, the Chern numbers of the positive-frequency bands are $C_n = -1, 2, 0, -1$, for $n = 1, 2, 3, 4$, shown in Fig.~\ref{fig:homog_plasma_spectrum}.  The Chern number of the zero-frequency band is 0, and the Chern numbers of the negative-frequency bands are the negative of their positive-frequency partner.  For $k_z < k_z^*$, the band structure smoothly transitions to $k_z = 0$.  At $k_z=0$, bands 2 and 4 become $X$ waves, band 3 is the $O$ wave, and band 1 has degenerate frequency $\w = 0$.  Different Chern numbers are obtained for $k_z > k_z^*$: $C_n = -1, 1, 1, -1$ for $n= 1, 2, 3, 4$.  Hence, multiple plasma bands are topologically nontrivial.  If the direction of $\v{B}_0$ is reversed, the Chern number also flips sign.  The gap Chern number is $\sum_{n=-4}^1 C_n = C_1 = -1$ for any $k_z$.

The gap between the first and second band of the plasma can overlap with a forbidden band in vacuum.  The dispersion relation for electromagnetic waves in vacuum is $\w^2 = c^2 k^2$.  For nonzero $k_z$, there is a forbidden region for $\w^2 < c^2 k_z^2$ in which waves cannot propagate.  If the plasma parameters can be engineered such that the band gaps in the plasma and vacuum overlap, bulk-boundary correspondence implies the existence of a undirectional surface mode crossing the gap. 

Previous work has considered a wave propagating at the planar, discontinuous interface of a semi-infinite, uniform-density magnetized plasma and vacuum \cite{yang:2016}.  However, a sharp interface is not physically realizable for gaseous plasma.  The interface width is typically limited by classical or turbulent diffusion processes and may be larger than the length scale of the wave.  A notable exception is nonneutral plasma, for which the interface width can be made comparable to the Debye length \cite{danielson:2006}.

\begin{figure}
		\includegraphics{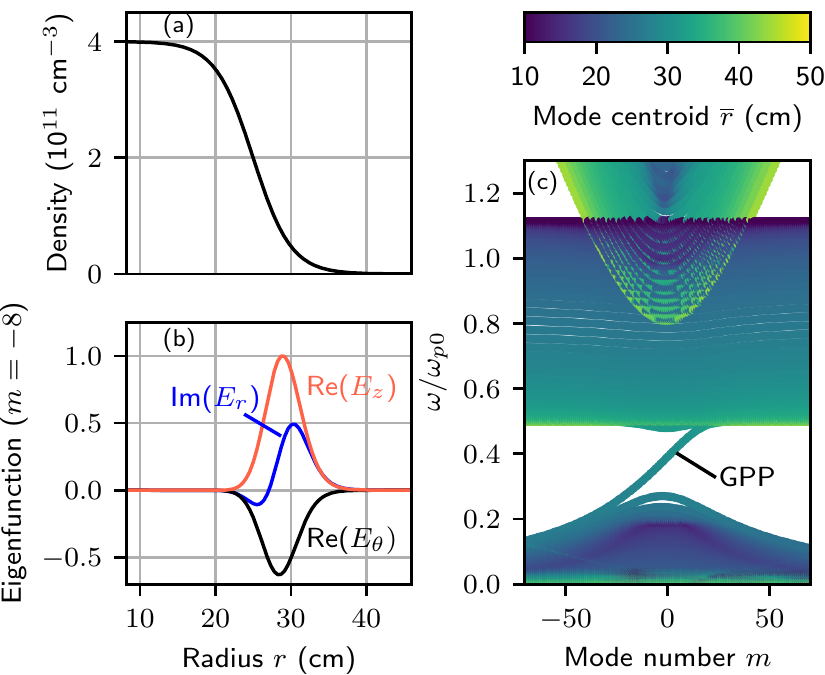}
		\caption{Spectrum of an inhomogeneous magnetized plasma.  Here, $ck_z/\w_{p0} = 0.8$ and $\s = 0.5$.  (a) Plasma density as a function of radius. (b) Nonzero components of GPP electric field at azimuthal mode number $m=-8$ .  (c) Spectrum as a function of $m$, where color corresponds to the mode centroid of the energy in the electric field.  The GPP dispersion relation is indicated and crosses the band gap.   There is another mode (not shown) in the numerical solution which is localized to the inner wall and stems from the artificial conducting-wall boundary condition.}
		\label{fig:inhomog_plasma_spectrum}
	\end{figure}

To determine whether the GPP can propagate in a realistic plasma, we consider a cylindrical plasma with magnetic field aligned along the $z$ axis.  We take into account a density profile that varies smoothly with radius, as shown in Fig.~\ref{fig:inhomog_plasma_spectrum}(a).  For simplicity, we assume a uniform magnetic field.  We assume the background plasma has azimuthal symmetry and translational symmetry in $z$.  The wave equation for an inhomogeneous cold plasma is simply \eqnref{homogeneous_cold_plasma_equations} with the replacement $n_e \to n_e(r)$.  For the density profile we use $n_e = \tfrac{1}{2} n_0 ( \tanh[ (r_0 - r)/L_n] + 1)$, where $L_n$ is the length scale over which the density decays.

We decompose eigenmodes as $f(\v{x}, t) = f(r) e^{i(m\th + k_z z - \w t)}$.  We solve the radial eigenvalue equation using the spectral code Dedalus \cite{burns:2020}.  Numerically, we consider a radial domain $[a, b]$ where $a > 0$, and for simplicity apply conducting-wall boundary conditions at both $r=a$ and $r=b$.  Since the mode of interest is a surface wave localized near $r=r_0$, a conducting boundary at $r=a$ can be used even when the physical situation has no inner wall as long as the surface wave has sufficiently small amplitude at $r=a$.  It would be preferable to use the more physical boundary condition of no inner wall and requiring only regularity at $r=0$, but we are restricted by our current numerical tools.  This more physical geometry could in principle allow the existence of another class of body modes~\cite{trivelpiece:1959}.
However, for the specific parameters used here, consideration of the dispersion relation near the plasma center indicates there can be no propagating modes.

The eigenmodes and spectrum are shown for one set of parameters in Figs.~\ref{fig:inhomog_plasma_spectrum}(b) and \ref{fig:inhomog_plasma_spectrum}(c).  Here, we take $r_0 = 25$ cm, $L_n = 5$ cm, $B_0 = 0.1$ T, and $n_0 = 4 \times 10^{11}$ cm$^{-3}$, which gives $\s = \W_e / \w_{p0}= 0.5$, where $\w_{p0}$ is the plasma frequency computed with $n_0$.  We take $ck_z / \w_{p0} = 0.8$.  These parameters have been chosen because they are accessible to existing laboratory devices.  For example, the Large Plasma Device (LAPD) has reported similar magnetic field values and density profiles \cite{gekelman:2016,maggs:2007}.  At these densities and typical electron temperatures ($\sim 10$~eV), the earlier assumptions justifying the cold plasma model are well satisfied.

In Fig.~\ref{fig:inhomog_plasma_spectrum}(c), the GPP crosses the band gap $0.3 < \w / \w_{p0} < 0.5$.  The electric-field polarization of the eigenfunction is displayed in Fig.~\ref{fig:inhomog_plasma_spectrum}(b) for $m=-8$, which shows that the GPP is a surface wave localized to the region between the plasma and vacuum.  There are no other waves for the GPP to scatter into at this $k_z$.

When $\w > \W_e$, upper hybrid modes become accessible.  These modes are identifiable as upper hybrid because their frequency is approximately independent of $m$ and the eigenfunctions are localized around the radial location corresponding with the frequency of local upper hybrid oscillations, $\w^2 = \w_p^2(r) + \W_e^2$.  The upper hybrid modes in the low- and intermediate-density region restrict the gapped frequency range and constitute a distinct difference from the spectrum of a plasma and vacuum separated by a sharp interface.  If the plasma density is uniform and discontinuously jumps to zero, the only upper hybrid frequency is $\w_{p0}^2 + \W_e^2$.

\begin{figure}
		\includegraphics{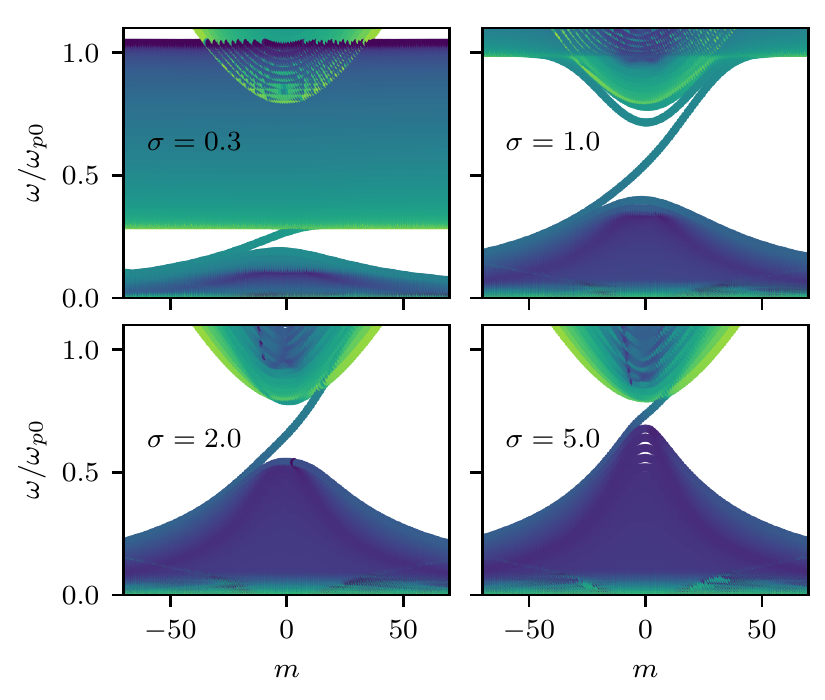}
		\caption{Spectrum for various magnetic field strengths $\s = |\Omega_e|/\w_{p0}$.  Other parameters and color scale are as in Fig.~\ref{fig:inhomog_plasma_spectrum}.}
		\label{fig:inhomog_spectrum_vary_sigma}
	\end{figure}

In a laboratory plasma, the strength of the applied magnetic field is one of the simplest parameters to adjust experimentally and is thus an important control knob.  Figure~\ref{fig:inhomog_spectrum_vary_sigma} shows how the spectrum varies with the magnetic field strength.  For $\W_e < \w_{p0}$, the band gap shrinks because the lowest upper hybrid frequency decreases, and for $|\W_e| > \w_{p0}$, the band gap shrinks because the top of the lower band rises to meet the bottom the upper band.  Maximizing the size of the bandg ap, which is achieved for $\W_e \approx \w_{p0}$, will isolate the GPP from other modes and ease its detection.

Our discussion has focused consideration on a single $k_z$.  The physical system itself is three dimensional, which unlike the two-dimensional case, is not expected to have topologically suppressed scattering \cite{ando:2013}.  However, the existence of the surface wave is still protected topologically against small perturbations.  Furthermore, magnetized plasmas tend to be much more uniform along the magnetic field than perpendicular to it.  Physically, this is because charged particles stream freely along a magnetic field line but are tied to the field line in the perpendicular direction by the cyclotron motion.  As a result, plasma nonuniformities along a magnetic field are smoothed out very quickly, and there would be very low spectral power in modes with $k_z / k_\perp \sim 1$.  This separation of scales has been observed to hold in the LAPD, with flutelike drift-Alfv\'{e}n perturbation modes that have very small $k_z$ \cite{penano:2000} and fluctuate slowly (tens of kHz) compared to the GPP ($\sim$GHz).  Therefore, to a first approximation, the system is translationally invariant and $k_z$ will be conserved. 

In summary, we have shown that the gaseous plasmon polariton, which arises from the nontrivial topology of waves in a bulk magnetized plasma, can exist at the plasma-vacuum interface with a realistic, gradual plasma density falloff.  The density scale length can be comparable to the wavelength.  For certain choices of plasma density and magnetic field, the wave propagates in a gapped frequency range and thus may be able to serve as a protected probe of plasma in tokamaks and other plasma devices.  We have shown that such parameter regimes are achievable in present-day cylindrical plasma devices, such as the Large Plasma Device at the Basic Plasma Science Facility.  Laboratory experiments to confirm the existence of this topological edge mode are therefore in reach.  Properties of the wave that can be predicted and compared with measurements include the frequency, dispersion relation, radial localization, and polarization.  Such experiments could confirm the first controlled observation of a wave of topological origin in a gaseous plasma.

\begin{acknowledgments}
We acknowledge useful discussions with Troy Carter, Bart Van Compernolle, George Morales, Hong Qin, and Shreekrishna Tripathi.  J.~B.~P.'s work was performed under the auspices of the U.S.\ Department of Energy by Lawrence Livermore National Laboratory under Contract No.\ DE-AC52-07NA27344.  J.~B.~P.\ and J.~B.~M.\ would like to acknowledge the workshop “Vorticity in the Universe” held at the Aspen Center for Physics in the summer of 2017 and supported by National Science Foundation Grant No.\ PHY-1607611, which played an important role in bringing about this work.  S.~M.~T.\ is supported by the European Research Council (ERC) under the European Union Horizon 2020 research and innovation program (Grant Agreement No.\ D5S-DLV-786780).  Z.~Z.\ is supported by the STC Center for Integrated Quantum Materials, NSF Grant No.\ DMR1231319.
\end{acknowledgments}

%

\clearpage
\onecolumngrid
\appendix
\setcounter{equation}{0}

\begin{center}
{\bf \large Supplementary Material for the paper \\
\emph{Topological Gaseous Plasmon Polariton in Realistic Plasma}}
\end{center}
\vspace{.4cm}

\section{Effective Hamiltonian for a cold plasma}
Here we derive the $9 \times 9$ matrix $H$ corresponding to the effective Hamiltonian for a cold plasma.  As given in Eq.~(1) of the main paper, the linearized equations for an infinite, homogeneous cold plasma are
	\begin{align*}
		\pd{\v{v}}{t} &= -\frac{e}{m_e} (\v{E} + \v{v} \times \v{B}_0), \\
		\pd{\v{E}}{t} &= c^2 \nabla \times \v{B} + \frac{e n_e}{\e_0} \v{v}, \\
		\pd{\v{B}}{t} &= -\nabla \times \v{E},
	\end{align*}
where electron dynamics are retained and ions are assumed stationary.  The applied magnetic field $\v{B}_0 = B_0 \unit{z}$ is uniform in space and constant in time.

We normalize time to $\w_p^{-1}$, length to $c/\w_p$, velocity to $e\ol{E} / m_e \w_p$, electric field to $\ol{E}$, and magnetic field to $\ol{E}/c$, where $\w_p = (n_e e^2/m_e \e_0)^{1/2}$ is the plasma frequency, $n_e$ is the background electron density, $e$ is the electron charge, $m_e$ is the electron mass, $\e_0$ is the permittivity of free space, $c$ is the speed of light, and $\ol{E}$ is some reference electric field.  In normalized variables, the linearized equations become
	\begin{subequations}
	\label{coldplasmaEOM_normalized}
	\begin{align}
		\pd{\v{v}}{t} &= - \v{E}  -\s \v{v} \times \unit{z}, \label{electronEOM} \\
		\pd{\v{E}}{t} &= \nabla \times \v{B} + \v{v}, \label{EfieldEOM} \\
		\pd{\v{B}}{t} &= -\nabla \times \v{E},
	\end{align}
	\end{subequations}
where $\s = \sign(B_0) \W_e / \w_{p}$, and $\W_e = |eB_0 / m_e|$ is the electron cyclotron frequency.  We Fourier transform in space and time, letting $\partial_t \to -i\w$ and $\nabla \to i\v{k}$.  We write the state vector in Cartesian components as
	\begin{equation}
		\ket{f} = \begin{bmatrix} v_x \\ v_y \\ v_z \\ E_x \\ E_y \\ E_z \\ B_x \\ B_y \\ B_z \end{bmatrix}
	\end{equation}
Then Eq.~\eref{coldplasmaEOM_normalized} can be rewritten as the matrix equation $\w \ket{f} = H \ket{f}$, where $H$ is the Hermitian $9 \times 9$ matrix,
	\begin{equation}
		H = \begin{bmatrix}
			0 & -i\s & 0 & -i & 0 & 0 & 0 & 0 & 0 \\
			i\s & 0 & 0 & 0 & -i & 0 & 0 & 0 & 0 \\
			0 & 0 & 0 & 0 & 0 & -i & 0 & 0 & 0 \\
			i & 0 & 0 & 0 & 0 & 0 & 0 & k_z & -k_y \\
			0 & i & 0 & 0 & 0 & 0 & -k_z & 0 & k_x \\
			0 & 0 & i & 0 & 0 & 0 & k_y & -k_x & 0 \\
			0 & 0 & 0 & 0 & -k_z & k_y & 0 & 0 & 0 \\
			0 & 0 & 0 & k_z & 0 & -k_x & 0 & 0 & 0 \\
			0 & 0 & 0 & -k_y & k_x & 0 & 0 & 0 & 0
		\end{bmatrix}.
		\label{Hamiltonian:no_regularization}
	\end{equation}
Hence, $H$ plays the role of the effective Hamiltonian.

\section{Regularization through plasma discreteness}
A regularization of the Hamiltonian is motivated based on the physical fact that at small enough scales, the plasma ceases to look like a continuous medium, i.e., at scales smaller than the average inter-particle spacing.  For electromagnetic waves at such small scales, the plasma cannot effectively respond.  The continuum fluid representation ignores this fact.

However, one can try to emulate the physical behavior within the fluid representation by modifying the Hamiltonian in the following way.  In the Fourier representation, we introduce a regularizing factor $r(\v{k})$ into the linearized equations of motion to suppress the plasma response at small scales.

Equation \eref{electronEOM} is modified to be
	\begin{equation}
		\partial_t \v{v} = -r(\v{k}) \v{E} - \s \v{v} \times \unit{z},
	\end{equation}
and Eq.~\eref{EfieldEOM} is modified to be
	\begin{equation}
		\partial_t \v{E} = \nabla \times \v{B} + r(\v{k}) \v{v}.
	\end{equation}
	
The regularized Hamiltonian matrix takes the form
	\begin{equation}
		H = \begin{bmatrix}
			0 & -i\s & 0 & -i r(\v{k}) & 0 & 0 & 0 & 0 & 0 \\
			i\s & 0 & 0 & 0 & -i r(\v{k}) & 0 & 0 & 0 & 0 \\
			0 & 0 & 0 & 0 & 0 & -i r(\v{k}) & 0 & 0 & 0 \\
			i r(\v{k}) & 0 & 0 & 0 & 0 & 0 & 0 & k_z & -k_y \\
			0 & i r(\v{k}) & 0 & 0 & 0 & 0 & -k_z & 0 & k_x \\
			0 & 0 & i r(\v{k}) & 0 & 0 & 0 & k_y & -k_x & 0 \\
			0 & 0 & 0 & 0 & -k_z & k_y & 0 & 0 & 0 \\
			0 & 0 & 0 & k_z & 0 & -k_x & 0 & 0 & 0 \\
			0 & 0 & 0 & -k_y & k_x & 0 & 0 & 0 & 0
		\end{bmatrix},
		\label{Hamiltonian:with_regularization}
	\end{equation}
which remains Hermitian.

Since the regularization is a mathematical convenience rather than an attempt at a faithful model of the actual discrete-particle nature of the plasma, the functional form of $r$ is a free choice.  For example, we can take
	\begin{equation}
		r(\v{k}) = \frac{1}{1 + (k_\perp/k_c)^p}
	\end{equation}
for some cutoff wavenumber $k_c$ and positive exponent $p$, where $k_\perp = \sqrt{k_x^2 + k_y^2}$.  Observe $r \approx 1$ for $k_\perp \ll k_c$, while $r \sim k_\perp^{-p}$ for $k_\perp \gg k_c$.

Mathematically, regularization is expected to restore smoothness in some sense at small spatial scales, i.e., wavenumbers at infinity.  This in turn should allow a compactification to the Riemann sphere by ensuring behavior at the north pole is sufficiently regular.  More technical details are given by Silveirinha [Silveirinha 2015]; compactification has also been discussed through odd viscosity [Tauber et al.~2019, Souslov et al.~2019].  Once compactification can be achieved, the Chern theorem, which states that for a compact two-dimensional manifold, the integral of the Berry curvature is an integer equal to the Chern number, is applicable.  The stereographic projection map to the Riemann sphere is never explicitly needed as the integration can be performed using coordinates on the plane.

\section{Chern number with and without regularization}
Without loss of generality, we take $k_z \ge 0$ and $\s > 0$.  As discussed in the main article, two bands touch at $k_z = k_z^*$ so there are two regimes, $0 < k_z < k_z^*$ and $k_z > k_z^*$, where $k_z^*$ (in normalized units) is determined by $(k_z^*)^2 = \s / (1 + \s)$.  We compute the Chern numbers $C_n$ by integrating the Berry curvature in the $\v{k}_\perp$ plane:
	\begin{equation}
		C_n = \frac{1}{2\pi} \int d\v{k}_\perp \, F_n(\v{k}_\perp)  =  \int_0^\infty d k_\perp\, k_\perp F_n(k_\perp),
		\label{perpendicular_Chern_integral}
	\end{equation}
where the second equality follows because this problem is isotropic in the plane perpendicular to $z$.

Without regularization, we find that $C_1$, $C_3$, and $C_4$ (corresponding to frequency bands 1, 3, and 4) are integer-valued, and $C_2$ (corresponding to band 2) is not an integer.  For the special case $k_z=0$, the Berry curvature was integrated by Hanson et al.~(2016) to be $C_2 = 1 + \s / \sqrt{1 + \s^2}$.  We find that this is in agreement with our numerical calculations for $k_z=0$ and also generalizes to nonzero $k_z$.  Our results from numerical integration are given in Table~\ref{tab:no_regularization}.

When we use regularization, we find that $C_2$ is now an integer (as it should be due to the Chern theorem) and the other bands have been left unchanged.  See Table~\ref{tab:regularization}.  The results are insensitive to the value of $k_c$ or $p$ used in the regularization.  E.g., we can use $k_c = 10$ or $k_c = 1000$ and obtain the same result.  We have used $p$ as large as $8$ and as small as $0.1$.  The numerical integration becomes more challenging at small $p$ because the Berry curvature decays slowly; we find numerically that at $k_\perp \gg k_c$, $k_\perp F(k_\perp) \sim k_\perp^{-(1+p)}$.  Based on our results, where we have used $p$ as small as 0.1, it appears that for any $p > 0$, the integral of the Berry curvature is an integer.

\begin{figure}
	\centering
	\includegraphics{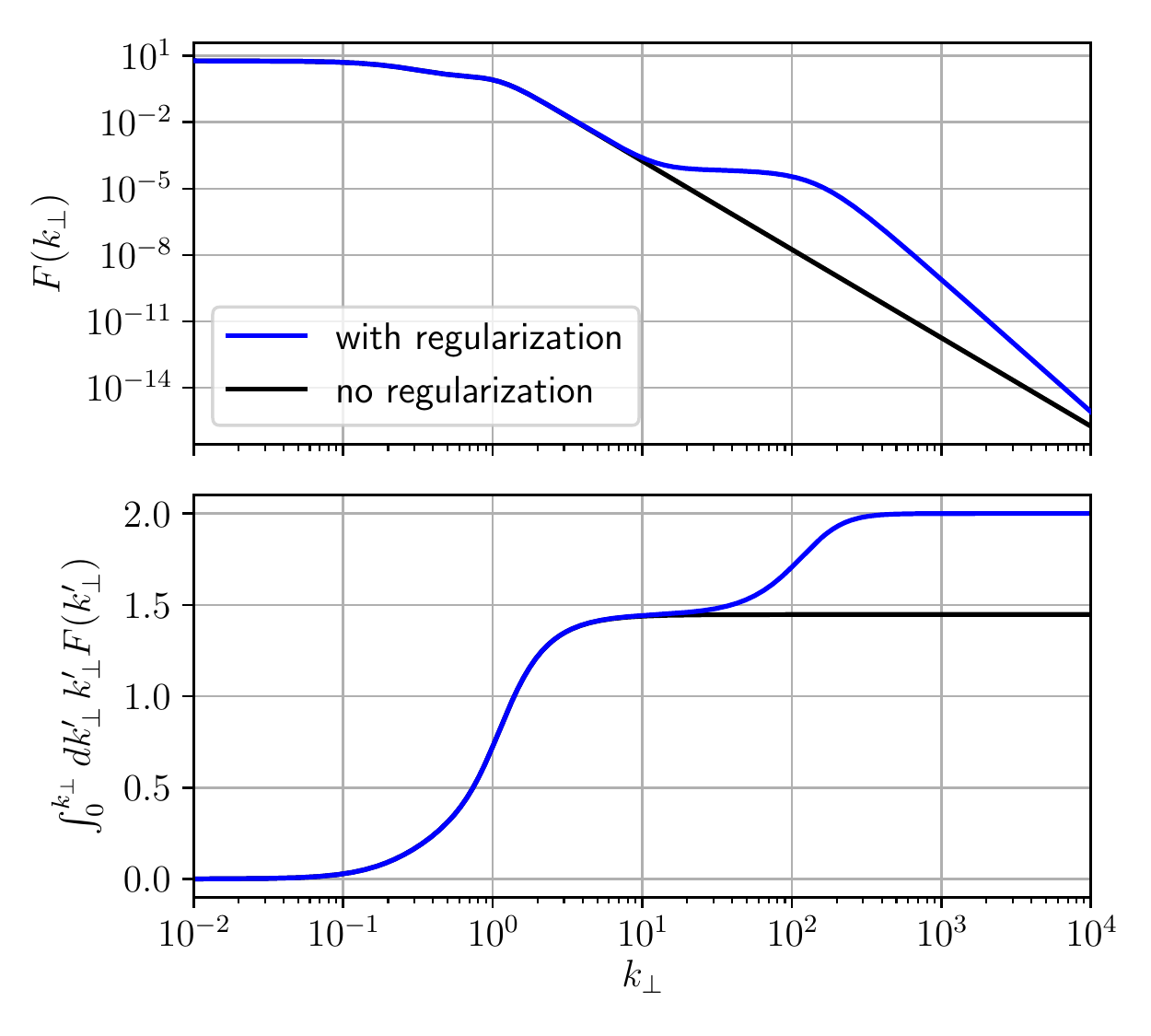}
	\caption{Top:  Berry curvature $F(k_\perp)$ for band 2, at $\s=0.5$ and $k_z = 0.4$, with and without regularization.  Bottom:  cumulative integral of the Berry curvature, with the Chern integral $C_2$ reproduced for $k_\perp \to \infty$ as in Eq.~\eref{perpendicular_Chern_integral}.  With regularization, the Berry curvature integrates to an integer, and without regularization it does not, in accordance with Tables \ref{tab:no_regularization} and \ref{tab:regularization}.  The regularization parameters here are $k_c = 100$ and $p=2$.}
	\label{fig:berrycurvature}
\end{figure}

Plots of the Berry curvature with and without regularization are shown in Figure~\ref{fig:berrycurvature}.

\setlength{\tabcolsep}{24pt}
\begin{table}
	\centering
	\caption{Chern number computed by integrating the Berry curvature in the $\v{k}_\perp$ plane without regularization, using the Hamiltonian of Eq.~\eref{Hamiltonian:no_regularization}.}
	\label{tab:no_regularization}
	\begin{tabular}{rcc} 
	\toprule
					& $k_z < k_z^*$  &  $k_z > k_z^*$ \\ \colrule
		$C_4$	&	$-1$						&		$-1$ \\
		$C_3$	&	$0$						&		$1$ \\[1ex]
		$C_2$	&	$1 + \dfrac{\s}{\sqrt{1 + \s^2}}$	&		$\dfrac{\s}{\sqrt{1 + \s^2}}$ \\[2ex]
		$C_1$	&	$-1$						&		$-1$ \\ \botrule
	\end{tabular}
\end{table}

\begin{table}
	\centering
	\caption{Same as Table~\ref{tab:no_regularization}, but now regularization is adopted using the Hamiltonian of Eq.~\eref{Hamiltonian:with_regularization}.}
	\label{tab:regularization}
	\begin{tabular}{rcc} 
	\toprule
					& $k_z < k_z^*$  &  $k_z > k_z^*$ \\ \colrule
		$C_4$	&	$-1$						&		$-1$ \\
		$C_3$	&	$0$						&		$1$ \\
		$C_2$	&	$2$						&		$1$ \\
		$C_1$	&	$-1$						&		$-1$ \\ \botrule
	\end{tabular}
\end{table}

\end{document}